\pdfoutput=1
\documentclass[a4paper]{article}

\usepackage{INTERSPEECH2020}
\usepackage{breqn}

\title{Homophone-based Label Smoothing in End-to-End Automatic Speech Recognition}
\name{Yi Zheng$^1$, Xianjie Yang$^1$, Xuyong Dang$^1$}
\address{
  $^1$UBTECH Robotics Inc, China}
\email{zhengyiuestc@gmail.com, yangxjzwd@163.com, dangxuyong@163.com}

\begin{document}

\maketitle
\begin{abstract}
A new label smoothing method that makes use of prior knowledge of a language at human level, homophone, is proposed in this paper for automatic speech recognition (ASR). Compared with its forerunners, the proposed method uses pronunciation knowledge of homophones in a more complex way. End-to-end ASR models that learn acoustic model and language model jointly and modelling units of characters are necessary conditions for this method. Experiments with hybrid CTC sequence-to-sequence model show that the new method can reduce character error rate (CER) by 0.4\% absolutely.
\end{abstract}

\section{Introduction}
An intuitive way to generate the distribution from training data is one-hot distribution in which only the class given by ground truth is assigned with probability of $1$, while all other classes are assigned with probability of $0$ \cite{Rethinking}. Training a network with one-hot distribution can cause over confidence in the network’s output distribution \cite{Rethinking}. Over confidence of a network indicates that the network’s structure is less smoothing and can do harm to its ability of generalization \cite{Rethinking}\cite{Regularizing-NN}.

One method to solve over confidence problem is to generate a new distribution by adding one-hot distribution with a prior distribution and to learn the network with the new distribution \cite{Rethinking}. This method is called label smoothing (LS). Another way to solve over confidence problem is output regularizers: a maximum entropy based confidence penalty \cite{Regularizing-NN}, to which we will not pay attention. It is shown that LS is equal to adding a KL divergence penalty term which penalizes the deviation of predicted label distribution from the prior distribution, to the negative log-likelihood loss function \cite{Rethinking}\cite{Regularizing-NN}.
Furthermore, distillation exploits the idea of training a small network with distribution given by a large network by assuming that larger networks have better generalization \cite{Distilling}. The problem is still that it is hard to train large networks when dataset is small.

We will concentrate on the LS methods in this paper. The uniform and unigram LS strategy have been introduced and implemented for ASR \cite{Rethinking}\cite{Regularizing-NN}\cite{ESPNET}\cite{Hybrid CTC attention}\cite{Joint CTC attention}. Uniform LS assigns same probability $1/K$ to all characters, where $K$ is number of characters used as modelling units. Unigram LS assigns the probability of a character by its character frequency in the training dataset.
Both LS methods use the prior knowledge of language at shallow level. To the best of our knowledge, no further prior knowledge of language have been used to generate the prior distribution.

Homophones exist in Chinese, Japanese and Korean. In this paper, we consider the application of label smoothing in ASR by making use of deeper prior knowledges: homophone.
%The problems of being without LS or with simple LS based on seq2seq ASR model will be analyzed in next chapter.
%, more specifically the hybrid CTC-attention model \cite{Hybrid CTC attention} \cite{Joint CTC attention}.
\subsection{The sequence-to-sequence ASR models}
Encoder-decoder based sequence-to-sequence (seq2seq) models with attention mechanism have shown state-of-the-art performance in ASR. A hybrid CTC-seq2seq model is introduced in \cite{Joint CTC attention}. In this hybrid structure, CTC together with seq2seq can enforce monotonic alignment between speech and label sequences. The discussion in this paper concerns on the seq2seq structure that is composed of RNN-like encoder and decoder. One critical feature of the seq2seq model is that it can combine acoustic model (AM) and language model (LM) together.

Denoting the variable length $T$ input frames as $\bm{x}=(x_1,\cdots, x_T)$, $U$ length output characters sequence as $\bm{c}=(c_1,\cdots, c_U)$, $c_u \in \{1,\cdots,K \}$, where $K$ is the number of characters, which are modelling units.

The seq2seq model predicts label distribution at position $u$ conditioning on the previous labels by the following recursive procedures \cite{Hybrid CTC attention}\cite{Joint CTC attention},
\begin{equation}
  p(\bm{c} | \bm{x}) = \prod_{u} p(c_u | \bm{x}, c_{1:u-1}).
  \label{seq2seq-probility}
\end{equation}
%$$
%  \bm{h} = \text{Encoder}(\bm{x})
%  \label{hidden-status}
%$$
%$$
%  c_{u} \sim \text{AttentionDecoder}(\bm{h},c_{1:u-1}).
%  \label{Attention-decoder}
%$$
With $c_u$, $s_{u-1}$ and $a_{u}$, the decoder can generate next label $c_u$ and update the status as following,
\begin{equation}
  c_u = \text{Generate}(a_u, s_{u-1}),
  \label{generate next character}
\end{equation}
\begin{equation}
  s_u = \text{Recurrency}(s_{u-1},a_u,c_u),
  \label{decoder-recurrency}
\end{equation}
where $s_{u-1}$ is the decoder status and $a_u$ is the content of input given by attention mechanism \cite{Hybrid CTC attention}\cite{Joint CTC attention}. Combining equation (\ref{generate next character}) and  (\ref{decoder-recurrency}), one obtains
\begin{equation}
  c_u = \text{Generate}(a_u, \text{Recurrency}(s_{u-2},a_{u-1},c_{u-1})).
  \label{generate decoder recurrency together}
\end{equation}
When the ground truth $\bm{c}^*$ is given, the loss function of the seq2seq model can be computed as
\begin{equation}
  \mathcal{L}(\theta,\bm{c}^*) \triangleq -\log p(\bm{c}^* | \bm{x})=- \sum_u \log p(c_u^* | \bm{x},c_{1:u-1}^*),
  \label{Seq2seq-loss}
\end{equation}
where $c_{1:u-1}^*$ is the ground truth of the previous characters and $\theta$ is the parameters of seq2seq model.

\subsection{Analysis to distribution problems of homophones}
To analyze the distribution of homophones generated from training corpus, let us consider two pairs of training data $(\bm{x}^i, \bm{c}^i)$, $i=1,2$, where $\bm{c}^i=(c^i_{1},\cdots,c^i_{U},c^i_{U+1})$, and $c^1_{u} = c^2_{u}$, for $u=1,\cdots,U$. We assume further that $c^1_{U+1}$ is a homophone of $c^2_{U+1}$, i.e., $c^1_{U+1}$ and $c^2_{U+1} $ have same pronunciation.

Given the fixed pair $(\bm{x}^1, \bm{c}^1)$, we consider equation (\ref{generate decoder recurrency together}) with respect to $c_{U+1}^1$ and $c_{U+1}^2$ respectively at position $u=U+1$, then obtain
\begin{equation}
  c^1_{U+1} = \text{Generate}(a^1_{U+1}, \text{Recurrency}(s^1_{U-1},a^1_{U},c^1_{U})),
  \label{generate decoder recurrency together sen1}
\end{equation}
\begin{equation}
  c^2_{U+1} = \text{Generate}(a^1_{U+1}, \text{Recurrency}(s^1_{U-1},a^1_{U},c^1_{U})).
  \label{generate decoder recurrency together sen2}
\end{equation}
It is easy to see that the following probability equation with respect to $c^1_{U+1}$ and $c^2_{U+1}$ holds
\begin{equation}
  p(c^1_{U+1} | \bm{x}^1, c^1_{1:U}) = p(c^2_{U+1} | \bm{x}^1, c^1_{1:U}),
  \label{probability approximate c1 and c2}
\end{equation}
since the input terms $a^1_{U+1}$, $a^1_U$ and $s_{U-1}^1$, on the right side of (\ref{generate decoder recurrency together sen1}) and (\ref{generate decoder recurrency together sen2})  are identical. Identity of $s_{U-1}^1$ in the two equations can be concluded from equation (\ref{decoder-recurrency}) and the assumption of $(\bm{x}^1,\bm{c}^1)$. Similar analysis to $(\bm{x}^2, \bm{c}^2)$ can result in
\begin{equation}
  p(c^1_{U+1} | \bm{x}^2, c^2_{1:U}) = p(c^2_{U+1} | \bm{x}^2, c^2_{1:U}).
  \label{probability approximate c1 and c2 in second pair}
\end{equation}

If one generates distribution from $(\bm{x}^1,\bm{c}^1)$ without LS or with simply LS, the assigned probability $p(c^1_{U+1}| \bm{x}^1, c^1_{1:U})$ is significantly higher than the assigned probability $p(c^2_{U+1}| \bm{x}^1, c^1_{1:U})$ (saying the probability of $c^2_{U+1}$ is not distinguishable from other characters that are not homophones of $c^1_{U+1}$) within training data pair $(\bm{x}^1, \bm{c}^1)$, i.e.,
\begin{equation}
  p(c^1_{U+1} | \bm{x}^1, c^1_{1:U}) \gg p(c^2_{U+1} | \bm{x}^1, c^1_{1:U}).
  \label{probability much greater}
\end{equation}
Similar analysis to $(\bm{x}^2, \bm{c}^2)$ results in
\begin{equation}
  p(c^1_{U+1} | \bm{x}^2, c^2_{1:U}) \ll p(c^2_{U+1} | \bm{x}^2, c^2_{1:U})
  \label{probability much smaller}.
\end{equation}
Equation (\ref{probability approximate c1 and c2}) and (\ref{probability much greater}) contradict with each other, equation (\ref{probability approximate c1 and c2 in second pair}) and (\ref{probability much smaller}) contradict with each other too.

Furthermore, considering that $c^1_{1:U}$ equals with $c^2_{1:U}$, the only difference between the left hand side of (\ref{probability much greater}) and that of (\ref{probability much smaller}) is the difference between $\bm{x}^1$ and $\bm{x}^2$. But the left hand side of (\ref{probability much greater}) and that of (\ref{probability much smaller}) are assigned with very different probability. This will cause the AM to overfit to the slight input features difference between $\bm{x}^1$ and $\bm{x}^2$, considering that $\bm{x}^1$ and $\bm{x}^2$ share most of phonic information, in form of FBANK features, which is useful for ASR ($\bm{c}^1$ and $\bm{c}^2$ share same pronunciation).
%The contradiction will further do harm to the networks training (especially acoustic model). Because the acoustic model is given different or even contradicted information by $\bm{c}^1$ and $\bm{c}^2$. The phenomenon will finally lead the acoustic model to overfit to the slight difference between $a^1_T$ and $a^2_T$

To alleviate this phenomenon, we propose a new LS strategy based on homophones to make the following equation holds
\begin{equation*}
  p(c^1_{U+1} | \bm{x}^1, c^1_{1:U}) \approx p(c^1_{U+1} | \bm{x}^2, c^2_{1:U}),
  \label{probability similarity be label smoothing}
\end{equation*}
for $(\bm{x}^i, \bm{c}^i)$, $i=1,2$. The new LS strategy can help to generate smoothing distribution for homophones in general cases.

%The no longer contradicted supervised training information can benefit first the training procedure by preventing the acoustic model from getting confused and overfitted, then improve generazation----in inference.
To sum up, two basic conditions are required to make homophone label smoothing work:

\begin{itemize}
\item The ASR model should be an end-to-end one that learns AM and LM jointly and simultaneously,
    %Espnet is one of the ASR models that satisfy this requirement \cite{Hybrid CTC attention}\cite{Joint CTC attention}. This hybrid model is a combination of seq2seq model and CTC loss together, in which CTC attached to encoder only. The seq2seq model models acoustic feature and language together, while the shared encoder attached with CTC model acoustic feature only. In this model structure, an additional RNN-based LM can be attached via so called shallow fusion [CITATION].
\item The modelling units should be characters, i.e., the modelling units should be at the level where the homophone phenomenon take place.
\end{itemize}

\section{Homophone-based label smoothing}
\subsection{Over confidence penalty by KL divergence}
%\subsubsection{KL divergence in ASR}
%As mentioned in paper \cite{Rethinking} \cite{Regularizing-NN}, a way to penalize over confidence is by adding a KL divergence term.
Let us first introduce label smoothing in ASR. For position $u$, instead of learning the one-hot distribution $v_{c^*_u}^{\text{OH}}(\bm{y}_u)$ in which $c^*_u$ is the ground truth at $u$, one can train the networks with the following label smoothing distribution
\begin{equation}
p^{\prime} (\bm{y}_u) = (1-\beta)v_{c^*_u}^{\text{OH}}(\bm{y}_u) + \beta v(\bm{y}_u),
\label{original label smoothing definition}
\end{equation}
where
\begin{equation}
  v_{c^*_u}^{\text{OH}} (y_{u,k}) \triangleq
  \begin{cases}
    1, &\text{if $k=k_0$,}\\
	0, &\text{others},\\
  \end{cases}
\label{one-hot distribution}
\end{equation}
is the one-hot distribution and $k_0$ is the index of $c^*_u$ in $\{1,\cdots,K\}$.

When the prior distribution $v$ is given and considering the position $u$, networks learning with smoothing distribution $p^{\prime}$ is equivalent to adding KL divergence between the distribution $v$ and the network’s output distribution $p$ to the negative log-likelihood \cite{Rethinking}\cite{Regularizing-NN}. It deserves to point out that the negative log-likelihood is cross entropy between one-hot distribution and networks' output distribution. It results in, together with the single term of $u$ in right hand side of equation (\ref{Seq2seq-loss}), a new loss function at position $u$ with respect to the training data pair $(\bm{x},\bm{c}^*)$
\begin{equation}
\begin{split}
\mathcal{L}(\theta, c_u^*)  = & - (1-\beta) \log p(c_u^* | \bm{x},c_{1:u-1}^*)\\
& + \beta D_{\text{KL}} ( v(\bm{y}_u) || p(\bm{y}_u | \bm{x},c_{1:u-1}^*) ), \\
\end{split}
\label{single original label smoothing penalty}
\end{equation}
in which we use notation $v(\bm{y}_u)$ to indicate that the distribution is not connected with $c_u^*$,~$\bm{y}_u = (y_{u,1},\cdots,y_{u,K})$ is the vector of random variables over all characters at position $u$. The parameter $\beta$ is for the trade off between the negative log-likelihood and KL divergence penalty term.

As discussed in last chapter, one can make use prior language knowledge, homophone, to generate distribution $v(\bm{y}_u)$ such that it depends on $c_u^*$, i.e. $v(\bm{y}_u) \triangleq v_{ c_u^*}(\bm{y}_u)$, in which the subscript $c_u^*$ indicates that the prior distribution $v_{ c_u^*}(\bm{y}_u)$ is related to $c_u^*$. With $v(\bm{y}_u) = v_{ c_u^*}(\bm{y}_u)$ holds, equation (\ref{single original label smoothing penalty}) becomes
\begin{equation*}
\begin{split}
\mathcal{L}_{\text{HOMO-LS}}(\theta, c_u^*) = & - (1- \beta) \log p(c_u^* | \bm{x},c_{1:u-1}^*) \\
& + \beta D_{\text{KL}}(v_{ c_u^*}(\bm{y}_u) || p(\bm{y}_u | \bm{x},c_{1:u-1}^*)). \\
\end{split}
\label{single homo label smoothing penalty}
\end{equation*}
Correspondingly, the loss function of seq2seq model with homophone-based prior distributions for sequence $\bm{c}^*$ is
\begin{equation*}
\begin{split}
& \mathcal{L}_{\text{HOMO-LS}}(\theta, \bm{c}_u^*) \\
%= & \sum_u \Big( - (1-\beta)  \log p(c_u^* | \bm{x},c_{1:u-1}^*) \\
% & + \beta D_{\text{KL}}(v_{ c_u^*}(\bm{y}_u) || p(\bm{y}_u | \bm{x},c_{1:u-1}^*)) \Big) \\
= & - (1-\beta) \sum_u \log p(c_u^* | \bm{x},c_{1:u-1}^*) \\
& + \beta \sum_u   D_{\text{KL}}(v_{ c_u^*}(\bm{y}_u) || p(\bm{y}_u | \bm{x},c_{1:u-1}^*)),
\end{split}
\label{sequence homo label smoothing penalty}
\end{equation*}
in which KL divergence is defined as
\begin{equation*}
\begin{split}
& D_{\text{KL}}(v_{ c_u^*}(\bm{y}_u) || p(\bm{y}_u |  \bm{x},c_{1:u-1}^*)) \\
= & -\sum_k v_{ c_u^*}(y_{u,k}) \log \frac{p(y_{u,k} |  \bm{x},c_{1:u-1}^*)}{v_{ c_u^*}(y_{u,k})}.
\end{split}
\label{KL divergence}
\end{equation*}
It will be described in next subchapter how to generate the prior distribution $v_{ c_u^*}(\bm{y}_u)$.

\subsection{Prior distribution by homophones}
\subsubsection{Homophone with unigram based prior distribution}
In application, considering that $c_u^*$ can either has homophones or not, the prior distribution $v_{ c_u^*}(\bm{y}_u)$ can be defined correspondingly as
%To generate $v_{ c_u^*}(\bm{y}_u)$ from two sources of information, homophone pronunciation and unigram, and combine them together, one obtains
\begin{equation}
  v_{ c_u^*}(\bm{y}_u) \triangleq
  \begin{cases}
    v^{\text{HOMO}}_{ c_u^*}(\bm{y}_u), &\text{if $c^*_u$ has homophone(s),}\\
	v^{\text{UNI}}(\bm{y}_u), &\text{if $c^*_u$ has no homophone(s)},\\
  \end{cases}
\label{synthetical label smoothing define}
\end{equation}
where $v^{\text{UNI}}(\bm{y}_u)$ is unigram distribution generated from training corpus.
\subsubsection{Homophone with N-gram  based prior distribution}
The homophone label smoothing idea can be further improved by taking more complex contextual information into consideration. In this way, the prior distribution can be defined as following
\begin{equation*}
  v_{ c_u^*}(\bm{y}_u)  \triangleq  \\
   \begin{cases}
    \begin{aligned}
    & v^{\text{HOMO}}_{ c_u^*} (\bm{y}_u) \\
   %  & (1-\gamma) v^{\text{NG}}_{ c_u^*}(\bm{y}_u),
    \end{aligned}
     & \text{if $c^*_u$ has homophone(s),}\\
	 v^{\text{NG}}_{ c_u^*}(\bm{y}_u),
     &\text{if $c^*_u$ has no homophone(s)},\\
  \end{cases}
\label{synthetical label smoothing define ngram}
\end{equation*}
where $v^{\text{NG}}_{c_u^*}(\bm{y}_{u})$ denotes the label smoothing distribution than can be defined by a N-gram LM, which can be trained with a lager text corpus individually. The meaning of the distribution $v^{\text{NG}}_{ c_u^*}(\bm{y}_u)$ is the distribution over all characters at position $u$ according to the previous $N-1$ characters. The distribution has nothing to do with $c_u^*$ itself, the subscript $c_u^*$ is only for the consistency of symbols.

To define $v^{\text{HOMO}}_{ c_u^*}(\bm{y}_u)$, we consider arbitrary position $u$ and its corresponding ground truth character $c_u^*$. Considering that $c_u^*$ has homophones, we denote the set of characters which are homophones of $c_u^*$ as $\text{Homo}(c_u^*)$, whose size is $N$. Denoting $k_0$ as the index of $c_u^*$ in the set $\{1,\cdots,K\}$ and $\text{Homo}(k_0)$ as the set of indexes of elements of $\text{Homo}(c_u^*)$ in set $\{1,\cdots,K\}$, the homophone-based distribution $v^{\text{HOMO}}_{ c_u^*}(\bm{y}_u)$, in which $\bm{y}_u = (y_{u,1},\cdots,y_{u,K})$, can be defined as
\begin{equation}
  v^{\text{HOMO}}_{ c_u^*}(y_{u,k}) \triangleq
  \begin{cases}
    0.6, &\text{if $k = k_0 $,}\\
	\frac{0.3}{N}, &\text{if $k \in$ \text{Homo}$(k_0)$,}\\
    \frac{0.1}{K-(N+1)}, &\text{Others.}
  \end{cases}
\label{homophone label smoothing define}
\end{equation}
One would notice that $\text{Homo}(c_u^*)$ is not only determined by character $c_u^*$ itself, but also by its pronunciation. The pronunciation of $c_u^*$ is determined by tokenization of the sentence, since $c_u^*$ could be a polyphone. To choose right pronunciation for $c_u^*$ and determine $\text{Homo}(c_u^*)$, one must make right tokenization first, when we take Chinese language into consideration.

%\subsubsection{Homophone-unigram label smoothing}
%If one wants to take contextual information into consideration, the more synthetical strategy is to combine homophone LS with unigram LS as the following
%\begin{equation}
%  v_{ c_u^*}(\bm{y}_u)  \triangleq
%  \begin{cases}
%  \begin{aligned}
%    \gamma & v^{\text{HOMO}}_{ c_u^*} (\bm{y}_u) + \\
%    & (1-\gamma) v^{\text{UNI}}_{ c_u^*}(\bm{y}_u),
%  \end{aligned}
%    &\text{if $c^*_u$ has homophones,}\\
%	v^{\text{UNI}}_{ c_u^*}(\bm{y}_u), &\text{if $c^*_u$ has no homophones}.\\
%  \end{cases}
%\label{synthetical label smoothing define unigram}
%\end{equation}

To sum up, the homophone-based label smoothing strategy contains two steps: 1. generating homophone-based prior distribution. 2. training neural networks with prior distribution.

\section{Experiments}
The toolbox Espnet \cite{ESPNET}, together with Chinese language, are used for experiments. The modelling units are 6763 Chinese characters selected from level \uppercase\expandafter{\romannumeral1} and \uppercase\expandafter{\romannumeral2} of GB2312, 54 English letters (uppercase and lowercase), $\langle \text{UNK} \rangle$ ~and $\langle \text{space} \rangle$.

A composite corpus, which includes AISHELL-2 and some other smaller corpus, is used for the experiments. The corpus is of 1499 hours and is of about 1.5 million utterances, in which the test set is 3.9 thousand utterances. The reason for the relatively smaller test dataset is that the decoding with non-streaming hybrid CTC-seq2seq model is slow.
\subsection{Settings}
The encoder consists of CNN and LSTMP \cite{LSTMP}. The LSTMP-related part of encoder consists of 3 layers with 1024 units each. The projection units of LSTMP is 1024.

The decoder is of 2 layer LSTM with 1024 units each.

The attention mechanism used is content and location aware one \cite{Attention}. For the attention, the number of attention convolution channels is 10, and the number of attention convolution filters is 100. The dimension of attention is 1024.

For training, the weight for CTC is $\lambda = 0.5$, weight of KL divergence is $\beta = 0.4$, dropout parameter is set to 0.5. It deservers to point out that it takes more epochs for homophone LS to reach its best model, comparing with the non-LS model. The non-LS model reaches its best model at 8th epoch, while homophone LS takes 23 epochs to reach its best model, under the same optimizer configuration. One can choose one-hot distribution at beginning of training to speed up convergence and apply homophone-based LS at end of training to guarantee better generalization \cite{Rethinking}, but this experiment is not conducted in this paper.

For decoding, the weight for CTC is $\lambda=0.6$, beam search size is set to 20. The decoding test is without shallow fusion LM. One can download the source code from ``https://github.com/zhengyiuestc'' to check for more details.

Several LS experiments are implemented with the same network configuration above: non-LS, unigram LS, homophone with unigram LS and homophone with bigram LS. The unigram distribution is generated by the text of training corpus. The N-gram LM is choosen to be the bigram LM, which is trained on the CLUECorpus2020 corpus with KenLM tool \cite{CLUECorpus2020}.

The results in Table \ref{tab:experiments results} indicates that homophone LS can reduce CER by $0.4\%$ absolutely comparing to its best competitor non-LS.
\begin{table}[th]
  \caption{CER results of different LS}
  \label{tab:experiments results}
  \centering
  \begin{tabular}{ l  c }
    \toprule
    \multicolumn{1}{l}{\textbf{Methods}} & \multicolumn{1}{l}{\textbf{CER test}} \\
    \midrule
    Non-LS                       & $7.9\%$~~~          \\
    Unigram LS                  & $9.5\%$~~~          \\
    Homophone with unigram LS                & $7.5\%$~~~          \\
    Homophone with bigram LS                & $8.8\%$~~~ \\
    %Homophone-unigram LS              & $8.4\%$~~~         \\
    \bottomrule
  \end{tabular}

\end{table}

%  \begin{tabular}{ l c  c }
%    \toprule
%    \multicolumn{1}{l}{\textbf{Methods}} &
%                                         \multicolumn{1}{l}{\textbf{CER validation}} &\multicolumn{1}{l}{\textbf{CER test}} \\
%    \midrule
%    No LS                       &???& $89.9\%$~~~          \\
%    Unigram LS                    &???& $9.5\%$~~~          \\
%    Homophone LS                  &???& $?\%$~~~          \\
%    Synthetical LS                &???& $8.4\%$~~~         \\
%    \bottomrule
%  \end{tabular}

\section{Conclusions}

The experiment results indicate that the performance of seq2seq ASR models can be improved if the human-level prior language knowledge, homophone, is injected via label smoothing. Thought distillation idea can use the knowledge of larger networks to teach the smaller networks, the lager networks can not necessarily learn the prior knowledge at human's level. More ASR models with different structures will not be conducted in this paper but left as further discussion in next chapter.

\section{Further discussion}
To consider the model structures, besides seq2seq structure, homophone LS can be applied in transformer-based and CTC-like ASR models.
Further more, we discuss the possibility of apply homophone LS in on-streaming seq2seq and on-streaming transformer ASR models.
With regarding to LM, we consider to introduce the N-gram term to replace the unigram term in (\ref{synthetical label smoothing define}) with the aim to make use of broader text information.
At the end, the influence of homophone LS to decoding paths selection is analyzed based on seq2seq models which are shallowly fused with recurrent neural network LMs (RNN-LMs).
\subsection{Other ASR model structures}
\subsubsection{Transformer-based ASR models}
Transformer-based ASR models have shown their state-of-the-art performance recently \cite{Transformer ASR}. Since the new idea of this paper has only impact on the training distribution, one can apply the same analysis to transformer-based ASR models and obtain similar results.
\subsubsection{CTC-like Model}
If one considers the CTC part of the hybrid CTC-seq2seq network, it is composed of the encoder and CTC loss function. Noticing that the encoder has been restricted by label smoothing strategy via seq2seq part of the hybrid network during multi-task training, the performance of CTC decoding model, which takes the same encoder together with WFST language model \cite{EESEN}, will be improved from the new label smoothing strategy.

\subsection{On-streaming ASR models}
The on-streaming hybrid CTC-seq2seq structure and on-steaming transformer structure can also benefit from the new idea \cite{Streaming seq2seq}\cite{Streaming Transformer}. Comparing with non-streaming models, the main difference in on-streaming models is the conditional distribution in equation (\ref{seq2seq-probility}). This says the condition dependency in (\ref{seq2seq-probility}) and its followed equations is no longer depends on the whole input sequence $\bm{x}$ but part of $\bm{x}$.

\subsection{LM shallow fusion}
\subsubsection{Paths selection with LM shallow fusion}
Homophone LS can benefit the paths selection in one-pass decoding where on-streaming ASR models are shallowly fused with RNN-LMs \cite{Fusion}, when one considers the on-streaming hybrid CTC-seq2seq model and the on-streaming transformer model \cite{Streaming seq2seq}\cite{Streaming Transformer}.

Considering the probability accumulation of AM and LM during one-pass decoding procedure where the AM is not trained with homophone LS, if a wrong homophone is chosen with high probability by seq2seq model, it will be hard for LM to correct the error by accumulating AM and LM scores together. Because the probability difference between right and wrong homophone is too high. On other hand, homophone LS can reduce the probability difference between right and wrong homophones, such that LM can correct the wrong path selection in decoding.

\subsection{Extension of homophone smoothing by fuzzy pronunciation}
Equation (\ref{homophone label smoothing define}) can be further improved by fuzzy pronunciation via Chinese Pinyin as follows:
\begin{equation*}
  v^{\text{HOMO}}_{ c_u^*}(y_{u,k}) \triangleq
  \begin{cases}
    0.6, &\text{if $k = k_0 $,}\\
	\frac{0.15}{N}, &\text{if $k \in$ \text{Homo}$(k_0)$,}\\
    \frac{0.15}{M}, &\text{if $k \in$ \text{Simi}$(k_0)$,}\\
    \frac{0.1}{K-(N+M+1)}, &\text{Others,}
  \end{cases}
\label{homophone label smoothing define extended}
\end{equation*}
in which $\text{Simi}(k_0)$ is the set of indexes of elements of $\text{Simi}(c_u^*)$ in $\{1,\dots,K\}$ and $M$ is the size of $\text{Simi}(k_0)$. $\text{Simi}(c_u^*)$ contains all characters that have fuzzy pronunciation with $c_u^*$. Formally, it can be defined as following
$$\text{Simi}(c_u^*)=\{c | c \text{ shares fuzzy pronunciation with } c_u^*\}.$$
The character $c$ shares fuzzy pronunciation with $c_u^*$ means that either $c$ and $c_u^*$ share similar consonant (e.g. 'z' and 'zh') or similar compound vowel (e.g. 'in' and 'ing').

\bibliographystyle{IEEEtran}
\bibliography{mybib}

% \begin{thebibliography}{9}
% \bibitem[1]{Davis80-COP}
%   S.\ B.\ Davis and P.\ Mermelstein,
%   ``Comparison of parametric representation for monosyllabic word recognition in continuously spoken sentences,''
%   \textit{IEEE Transactions on Acoustics, Speech and Signal Processing}, vol.~28, no.~4, pp.~357--366, 1980.
% \bibitem[2]{Rabiner89-ATO}
%   L.\ R.\ Rabiner,
%   ``A tutorial on hidden Markov models and selected applications in speech recognition,''
%   \textit{Proceedings of the IEEE}, vol.~77, no.~2, pp.~257-286, 1989.
% \bibitem[3]{Hastie09-TEO}
%   T.\ Hastie, R.\ Tibshirani, and J.\ Friedman,
%   \textit{The Elements of Statistical Learning -- Data Mining, Inference, and Prediction}.
%   New York: Springer, 2009.
% \bibitem[4]{YourName17-XXX}
%   F.\ Lastname1, F.\ Lastname2, and F.\ Lastname3,
%   ``Title of your INTERSPEECH 2020 publication,''
%   in \textit{Interspeech 2020 -- 20\textsuperscript{th} Annual Conference of the International Speech Communication Association, September 15-19, Graz, Austria, Proceedings, Proceedings}, 2020, pp.~100--104.
% \end{thebibliography}

\end{document}